\documentclass{llncs}
\usepackage{lmodern}
\usepackage[T1]{fontenc}
\usepackage[utf8]{inputenc} %utf8x
\usepackage[normalem]{ulem}
\usepackage{amsmath,amssymb,color,bussproofs,epsfig,graphicx,verbatim,bm,galois,xcolor,xspace,todonotes,turnstile}
\usepackage{microtype}
\usepackage[pdftex,colorlinks=true,bookmarks=false,linkcolor=blue,citecolor=blue,urlcolor=blue] {hyperref}
\usepackage{multirow}
\usepackage{hhline}
\usepackage{etex}
\usepackage[all]{xy}
\usepackage{galois}
\usepackage{color}
\usepackage{subcaption}
\usepackage{galois}
\usepackage{tikz}
\usetikzlibrary{arrows,automata}

\newcommand{\AP}{{\textit{AP}}\xspace}

\newcommand{\trans}{\textit{trans}\xspace}
\newcommand{\FeatExp}{\textit{FeatExp}}
\newcommand{\true}{\textit{true}}
\newcommand{\false}{\textit{false}}

\newcommand{\U}{\textsf{U}}

\newcommand{\circled}[1]{\raisebox{.5pt}{\textcircled{\raisebox{-.9pt} {#1}}}}

\newcommand{\vending}{\textsc{VendingMachine}}

\newcommand{\sbr}[1]{\lbrack \! \lbrack #1 \rbrack \! \rbrack}

\newcommand{\poset}[2]{\ensuremath{\langle{#1},{#2}\rangle}}

\newcommand{\smv}{\textsc{NuSMV}}
\newcommand{\fsmv}{f\textsc{NuSMV}}

\newcommand{\elevator}{\textsc{Elevator}}

\newcommand{\Ff}{\mathbb{F}}
\newcommand{\Fff}{\mathcal{F}}
\newcommand{\Mmm}{\mathcal{M}}
\newcommand{\Mff}{\mathcal{MF}}
\newcommand{\Ttt}{\mathcal{T}}
\newcommand{\Kk}{\mathbb{K}}

\newcommand{\joinasym}{\ensuremath{\bm{\alpha}^{\textnormal{\textrm{join}}}}}
\newcommand{\joingsym}{\ensuremath{\bm{\gamma}^{\textnormal{\textrm{join}}}}}

\newcommand{\fignoreasym}{\ensuremath{\bm{\alpha}^{\textnormal{\textrm{fignore}}}}}
\newcommand{\fignoregsym}{\ensuremath{\bm{\gamma}^{\textnormal{\textrm{fignore}}}}}
\newcommand{\fignorea}[1]{\ensuremath{\fignoreasym_{#1}}}
\newcommand{\fignoreg}[1]{\ensuremath{\fignoregsym_{#1}}}

\definecolor{redgray}{rgb}{0.8,0.2,0.2}
\definecolor{greengray}{rgb}{0.2,0.6,0.2}
\definecolor{lightgray}{rgb}{0.8,0.8,0.8}
\definecolor{darkgreen}{rgb}{0.0,0.5,0.0}

\newcommand{\fs}{\textcolor{darkgreen}{\ensuremath{s}}}
\newcommand{\ft}{\textcolor{red}{\ensuremath{t}}}
\newcommand{\ff}{\textcolor{blue}{\ensuremath{f}}}
\newcommand{\fc}{\textcolor{brown}{\ensuremath{c}}}
\newcommand{\fv}{\textcolor{black}{\ensuremath{v}}}

\newcommand{\TR}[1]{%
  \raisebox{-.3mm}{\ensuremath{\xrightarrow{\smash{\!#1}}}}}
\newcommand{\ST}[1]{%
  \ensuremath{\circled{\tiny\strut \raisebox{.6ex}{#1}}}}

\title{Abstract Family-based Model Checking using Modal Featured Transition Systems: Preservation of CTL$^{\star}$ (Extended Version)
}

\author{ Aleksandar S. Dimovski
}

\institute{Faculty of Informatics, Mother Teresa University, Skopje, Mkd
}

\begin{document}
\maketitle

\begin{abstract}

  Variational systems allow effective building of many custom variants by using features (configuration options)
   to mark the variable functionality.
  In many of the applications, their quality assurance and formal verification are of paramount importance.
  Family-based model checking allows simultaneous verification of all variants
  of a variational system in a single run by exploiting the commonalities between the variants.
  Yet, its computational cost still greatly depends on the
  number of variants (often huge).

  In this work, we show how to achieve efficient family-based model checking of CTL$^{\star}$ temporal properties using
  variability abstractions and off-the-shelf (single-system) tools.
  We use variability abstractions for deriving abstract family-based model checking,
  where the variability model of a variational system is replaced with an abstract
  (smaller) version of it, called \emph{modal featured transition system}, which preserves the satisfaction of
  both universal and existential temporal properties, as expressible in CTL$^{\star}$.
  Modal featured transition systems contain two kinds of transitions, termed may and must transitions, which are defined by the conservative (over-approximating) abstractions
  and their dual (under-approximating) abstractions, respectively.
  The variability abstractions can be combined with different partitionings of the set
  of variants to infer suitable divide-and-conquer verification plans for the variational system.
  We illustrate the practicality of this approach for several
  variational systems.

\end{abstract}

\section{Introduction}\label{sec:introduction}
%\section{Introduction}

\noindent
Variational systems appear in many application areas
and for many reasons.
Efficient methods to achieve customization, such as \emph{Software Product Line Engineering} (SPLE) \cite{pl-patterns-book}, use \emph{features} (configuration options) to control presence and absence of the variable functionality \cite{fosd-book}. %,foda}.
Family members, called \emph{variants} of a \emph{variational system}, are specified in terms of features selected for that particular variant.  The reuse of code common to multiple variants is maximized.
The SPLE method is particularly popular in the embedded and critical system domain (e.g.\ cars, phones). In these domains, a rigorous verification and analysis is very important.  Among the methods included in current practices, \emph{model checking}\,\cite{katoen-beier} is a well-studied %formal
 technique used to establish %or refute
 that temporal logic properties hold for a system.

Variability and SPLE are major enablers, but also a source of complexity.
Obviously, the size of the configuration space (number of variants) is the limiting factor
to the feasibility of any verification technique.
 Exponentially many variants can be derived from few configuration options.
This problem is referred to as \emph{the configuration space explosion} problem.
%  However the model checking technology has not been designed to deal with configurable systems.
A simple ``brute-force'' application of a
single-system model checker to each variant is infeasible for realistic variational systems, due to the sheer number of variants.
This is very ineffective also because the same execution behavior is checked multiple times, whenever it is shared by some variants.
Another, more efficient, verification technique \cite{model-checking-spls,DBLP:journals/sttt/ClassenCHLS12} is based on
 using compact representations for modelling variational systems, which incorporate the commonality within the family.
We will call these representations variability models (or featured transition systems).
Each behavior in a variability model is associated with the set of variants able to produce it.  A specialized family-based model checking algorithm executed on such a model, checks an execution behavior only once regardless of how many variants include it.  These algorithms model check all variants simultaneously in a single run and pinpoint the variants that violate properties.
Unfortunately, their performance \emph{still} heavily depends on the size and complexity of the configuration space of the analyzed variational system.
Moreover, maintaining specialized family-based tools is also an expensive task.

In order to address these challenges, we propose to use standard, single-system model checkers  with an alternative, externalized way to combat the configuration space explosion.  We apply the so-called \emph{variability abstractions} to
a variability model which is too large to handle (``configuration space explosion''),
producing a more \emph{abstract model}, which is smaller than the original one.
We abstract from certain aspects of the configuration space, so that many of the configurations (variants) become
indistinguishable and can be collapsed into a single abstract configuration.
The abstract model is constructed in such a way that if some property holds for this abstract model it will also hold for the concrete model.
Our technique extends the scope of existing over-approximating variability abstractions \cite{spin15,fase17} which
currently support the verification of universal properties only (LTL and $\forall$CTL).
Here we construct abstract variability models which can be used to check arbitrary formulae of CTL$^{\star}$, thus including
arbitrary nested path quantifiers.
We use modal featured transition systems (MFTSs) for representing abstract variability models.
MFTSs are featured transition systems (FTSs) with two kinds of transitions, \emph{must} and \emph{may},
expressing behaviours that necessarily occur (must) or possibly occur (may).
We use the standard conservative (over-approximating) abstractions to define may transitions, and their dual
(under-approximating) abstractions to define must transitions. Therefore, MFTSs perform both
over- and under-approximation, admitting both universal and existential properties to be deduced.
Since MFTSs preserve all CTL$^{\star}$ properties, we can verify any such properties on the concrete
variability model (which is given as an FTSs) by verifying these on an abstract MFTS.
Any model checking problem on modal transitions systems (resp., MFTSs) can be reduced to two traditional model checking problems
on standard transition systems (resp., FTSs).
The overall technique relies on partitioning and abstracting concrete FTSs, until the point we obtain models with so limited variability (or, no variability) that it is feasible to complete their model checking in the brute-force fashion using the standard single-system model checkers.
 %In the extreme case, all variability is abstracted away, and the classic model checkers can be used to show some interesting properties for the entire system family.  This way we both lower the maintenance cost of family-based model checking tools (replacing a specialized model checker with a simpler meta-algorithm script), and we increase their performance/scalability to larger models (replacing a specialized model checker with a highly optimized standard model checker).
Compared to the family-based model checking, experiments show that the proposed technique %combined with the single-system model checker
achieves significant performance gains . %\snip\ on smaller benchmarks.
%On larger models the procedure (with \spin) scales better than \snip.

% to construct a divide-and-conquer
%verification plan that relies on using an existing single-system model checker. 

\section{Background} \label{sec:background}
%\section{Modelling SPLs}
\begin{comment}
We now introduce featured transition systems (FTSs) \cite{model-checking-spls} for modelling the behaviour
of all instances of variational systems.
Then, we describe fLTL temporal formulae \cite{model-checking-spls} for specifying properties of variational systems.
Finally, we define variability abstractions and the induced abstract FTSs introduced in \cite{spin15,sttt16}, which can be used
for more efficient verification of temporal properties.
\end{comment}

In this section, we present the background used in later developments.
%we begin with the definitions of featured transition systems (FTSs)
%and modal featured transition systems (MFTSs),  which are used as basic modeling formalisms of variational systems.  Then, we present
% CTL$^\star$ temporal formulae used to specify system properties.

\paragraph{\textbf{Modal Featured Transition Systems.}}
Let $\Ff = \{A_1, \ldots, A_n\}$ be a finite set of Boolean variables representing
the features available in a variational system.
A specific subset of features, $k \subseteq \Ff$, known as \emph{configuration}, specifies
a \emph{variant} (valid product) of a variational system.
We assume that only a subset \(\Kk \subseteq 2^{\Ff}\) of configurations are \emph{valid}.
An alternative representation of configurations is based upon
propositional formulae. Each configuration $k \in \Kk$ can be represented by a formula:
$k(A_1) \land \ldots \land k(A_n)$,
where $k(A_i) = A_i$ if $A_i \in k$, and $k(A_i) = \neg A_i$ if $A_i \notin k$ for $1 \leq i \leq n$.
%Since any configuration expressed as formula can be bijectively translated into a set of features,
We will use both representations interchangeably.
%The set of valid configurations is typically described by a feature model \cite{foda},
%but in this work we disregard syntactic representations of the set $\Kk$.

We recall the basic definition of a transition system (TS) and a modal transition system (MTS)
that we will use to describe behaviors of single-systems.
\begin{definition}
A transition system (TS) is a tuple $\Ttt=(S,Act,trans,I,AP,L)$, where $S$ is a set of states;
$Act$ is a set of actions; $trans \subseteq S \times Act \times S$ is a transition
relation;
$I \subseteq S$ is a set of initial states; $AP$ is a set of atomic propositions;
and $L : S \to 2^{AP}$ is a labelling function specifying which propositions hold in a state.  We write $s_1 \TR {~\lambda~} s_2$ whenever \((s_1,\lambda,s_2) \in \trans\).
\end{definition}
An \emph{execution} (behaviour) of a TS $\Ttt$ is an \emph{infinite} sequence $\rho = s_0 \lambda_1 s_1 \lambda_2 \ldots$
with $s_0 \in I$ such that $s_i \stackrel{\lambda_{i+1}}{\longrightarrow} s_{i+1}$ for all $i \geq 0$.
The \emph{semantics} of the TS $\Ttt$, denoted as $\sbr{\Ttt}_{TS}$, is the set of its executions.

MTSs \cite{DBLP:conf/lics/LarsenT88} are a generalization of transition systems that allows describing not just a sum of all behaviors of a system but also an over- and under-approximation of the system's behaviors. An MTS is a TS equipped with two transition relations: \emph{must} and \emph{may}. The former (must) is used to specify the required behavior, while the latter (may) to specify the allowed behavior of a system.
\begin{definition}
A modal transition system (MTS) is represented by a tuple $\Mmm=(S,Act,trans^{may},trans^{must},I,AP,L)$, where
$trans^{may} \subseteq S \times Act \times S$ describe may transitions of $\Mmm$;
$trans^{must} \subseteq S \times Act \times S$ describe must transitions of $\Mmm$,
such that $trans^{must} \subseteq trans^{may}$.
\end{definition}
The intuition behind the inclusion $trans^{must} \subseteq trans^{may}$ is that transitions that
are necessarily true ($trans^{must}$) are also possibly true ($trans^{may}$).
A \emph{may-execution} in $\Mmm$ is an execution with all its transitions in $trans^{may}$; whereas
a \emph{must-execution} in $\Mmm$ is an execution with all its transitions in $trans^{must}$.
We use $\sbr{\Mmm}_{MTS}^{may}$ to denote the set of all may-executions in $\Mmm$, whereas
$\sbr{\Mmm}_{MTS}^{must}$ to denote the set of all must-executions in $\Mmm$.

An FTS describes behavior of a whole family of systems in a \emph{superimposed} manner.  This means that it combines models of many variants in a single monolithic description, where the transitions are guarded by a \emph{presence condition} that identifies the variants they belong to.  The presence conditions $\psi$ are drawn from the set of feature expressions, $\FeatExp(\Ff)$, which are  propositional logic formulae over $\Ff$:
$  \psi ::= \true \mid A \in \Ff \mid \neg \psi \mid \psi_1 \land \psi_2$.
The presence condition $\psi$ of a transition specifies the variants in which the transition is enabled.  We write $\sbr{\psi}$ to denote the set of variants from $\Kk$ that satisfy $\psi$, i.e.\ $k \in \sbr{\psi}$ iff $k \models \psi$.
 %where $\models$ is the standard satisfaction relation of propositional logic.  For example, given \(\Ff=\{A, B\}\) with all four possible variants being valid, we get: \(\sbr{A \lor B} = \{ A \land B, A \land \neg B, \neg A \land B \}\).

%The combined behaviour of a whole system family is compactly represented with
%\emph{featured transition systems} \cite{model-checking-spls}. They are TSs where transitions are also labelled with
%feature expressions, $FeatExp(\Ff)$, which represent propositional logic formulae defined over $\Ff$
%as: $\psi ::= \true \mid A \in \Ff \mid \neg \psi \mid \psi_1 \land \psi_2$.
%The feature expression $\psi \in FeatExp(\Ff)$
%indicates for which variants the corresponding transition is enabled.

\begin{definition}
A featured transition system (FTS) represents a tuple $\Fff=(S,Act,trans,I,AP,L,\Ff,\Kk,\delta)$, where $S, Act, trans, I, AP$, and $L$
are defined as in TS; $\Ff$ is the set of available features; $\Kk$ is a set of valid configurations; and
$\delta: trans \to FeatExp(\Ff)$ is a total function decorating transitions with presence conditions (feature expressions).
\end{definition}
The \emph{projection} of an FTS $\Fff$ to a variant $k \in \Kk$, denoted as $\pi_k(\Fff)$, is the TS
$(S,Act,trans',I,AP,L)$, where $trans'=\{ t \in trans \mid k \models \delta(t) \}$.
We lift the definition of \emph{projection} to sets of configurations \(\Kk' \!\subseteq\! \Kk\),
 denoted as $\pi_{\Kk'}(\Fff)$, by keeping the transitions admitted by at least one of the configurations in $\Kk'$.
That is, $\pi_{\Kk'}(\Fff)$, is the FTS
$(S,Act,trans',I,AP,L,\Ff,\Kk',\delta)$, where $trans'=\{ t \in trans \mid \exists k \in \Kk'. k \models \delta(t) \}$.
The \emph{semantics} of an FTS $\Fff$, denoted as $\sbr{\Fff}_{FTS}$, is the union of behaviours
of the projections on all valid variants $k \in \Kk$, i.e.\ $\sbr{\Fff}_{FTS} = \cup_{k \in \Kk} \sbr{\pi_k(\Fff)}_{TS}$.

We will use modal featured transition systems (MFTS) for representing abstractions of FTSs.
MFTSs are variability-aware extension of MTSs.
%MTSs are a generalization of transition systems that allows describing not just a sum of all behavior of a system but also an over- and under-approximation of the behavior. A MTS is a transition system equipped with two transition relations: \emph{must} and \emph{may}. The former (must) is used to specify the required behavior of a system, while the latter (may) is used to specify the allowed behavior of a system.
\begin{definition}
A modal featured transition system (MFTS) represents a tuple $\Mff=(S,Act,trans^{may},trans^{must},I,AP,L,\Ff,\Kk,\delta^{may},\delta^{must})$, where
$trans^{may}$ and $\delta^{may}: trans^{may} \to FeatExp(\Ff)$ describe may transitions of $\Mff$;
$trans^{must}$ and $\delta^{must}: trans^{must} \to FeatExp(\Ff)$ describe must transitions of $\Mff$.
%such that $trans^{must} \subseteq trans^{may}$.
\end{definition}
The \emph{projection} of an MFTS $\Mff$ to a variant $k \in \Kk$, denoted as $\pi_k(\Mff)$, is the MTS
$(S,Act,trans'^{may},trans'^{must},I,AP,L)$, where $trans'^{may}=\{ t \!\in\! trans^{may} \mid k \!\models\! \delta^{may}(t) \}$, $trans'^{must}=\{ t \!\in\! trans^{must} \mid k \!\models\! \delta^{must}(t) \}$.
We define $\sbr{\Mff}^{may}_{MFTS}=\cup_{k \in \Kk} \sbr{\pi_k(\Mff)}^{may}_{MTS}$, and
$\sbr{\Mff}^{must}_{MFTS}=\cup_{k \in \Kk} \sbr{\pi_k(\Mff)}^{must}_{MTS}$.

\begin{figure}[t]
\centering
\begin{minipage}[b]{.59\textwidth}
\centering
\begin{tikzpicture}[->,>=stealth',shorten >=0.9pt,auto,node distance=1.2cm, semithick]
  \tikzstyle{every state}=[minimum size=.2pt,initial text={{\scriptsize start}}]

  \node[initial below,state] (A)     {\tiny 1};
  \node[state]         (B) [right of=A] {\tiny 2};
  \node[state]         (C) [right of=B] {\tiny 3};
  \node[state]         (D) [above right of=A]  {\tiny 4};
  \node[state]         (E) [above right of=C]  {\tiny 5};
  \node[state]         (F) [below right=4mm of C]  {\tiny 6};
  \node[state]         (G) [below right of=E]  {\tiny 7};
  \node[state]         (H) [right=11mm of G]  {\tiny 8};
  %\node[state]         (I) [right of=H]  {\tiny 9};

  \path[font=\tiny] (A) edge node {$\textit{pay} /\!v$} (B)
        (B) edge node {$\textit{change} /\!v$} (C)
        (G) edge node {$\textit{open} /\!v$} (H)
        (H) edge [in=-70,out=-25]  node[above,sloped] {$\textit{take} /\!v$} (A);

  \path[draw=darkgreen,fill=darkgreen,font=\tiny]
    (C) edge [bend left] node[above,sloped]
    {{\color{darkgreen}$\textit{soda} /\!s$}} (E)
    (E) edge [bend left] node[above,sloped]
    {{\color{darkgreen}$\textit{serveSoda} /\!s$}} (G);

  \path[draw=red,fill=red,font=\tiny]      (C) edge [bend right] node[below,sloped] {{\color{red}$\textit{tea} / t$}} (F)
        (F) edge [bend right,in=-180] node[above,sloped]
        {{\color{red}$\textit{~~serveTea} /\! t$}} (G);

  \path[draw=blue,fill=blue,font=\tiny]      (A) edge [bend right]
  node[below,sloped] {{\color{blue}$\textit{free} /\! f$}} (C)
        (G) edge [bend left=56,in=128,out=128]  node[above,sloped]
        {{\color{blue}$\textit{take} /\!f$}} (A);

  \path[draw=brown,fill=brown,font=\tiny]
        (G) edge [in=68,out=128] node[below,sloped] {{\color{brown}$\textit{take} / c$}} (C)
        (C) edge [bend right] node[above,sloped] {{\color{brown}$\textit{cancel} / c$}} (D)
        (D) edge [bend right] node[above,sloped] {{\color{brown}$\textit{return} /\! c$}} (A);

\end{tikzpicture}
\vspace{-10mm}
\caption{The FTS for \vending.}
\label{fig:FTS}
\end{minipage}%
\begin{minipage}[b]{.41\textwidth}
\centering
\begin{tikzpicture}[->,>=stealth',shorten >=0.8pt,auto,node distance=0.95cm,
                    semithick]
  \tikzstyle{every state}=[minimum size=.15pt,initial text={{\scriptsize start}}]

  \node[initial above,state] (A)     {\tiny 1};
  \node[state]         (B) [right of=A] {\tiny 2};
  \node[state]         (C) [right of=B] {\tiny 3};
  %\node[state]         (D) [above right of=A]  {\tiny 4};
  \node[state]         (E) [above right of=C]  {\tiny 5};
  %\node[state]         (F) [below right of=C]  {\tiny 6};
  \node[state]         (G) [below right of=E]  {\tiny 7};
  \node[state]         (H) [right of=G]  {\tiny 8};
  %\node[state]         (I) [right of=H]  {\tiny 9};

  \path[font=\tiny] (A) edge              node {$\textit{pay}$} (B)
        (B) edge              node {$\textit{change}$} (C)
        (G) edge   node {$\textit{open}$} (H)
        (H) edge   [bend left=20]  node[below,sloped] {$\textit{take}$} (A);

  \path[font=\tiny]      (C) edge [bend left] node[above,sloped] {{$\textit{soda}$}} (E)
        (E) edge [bend left] node[above,sloped] {{$\textit{serveSoda}$}} (G);

  %\path[draw=red,font=\tiny]      (C) edge [bend right] node[below,sloped] {{\color{red}$\textit{tea} / t$}} (F)
   %     (F) edge [bend right] node[above left] {{\color{red}$\textit{serveTea} / t$}} (G);

  %\path[draw=blue,font=\tiny]      (A) edge [bend right] node[below,sloped] {{\color{blue}$\textit{free} / f$}} (C)
   %     (G) edge [bend left=75]  node[above,sloped] {{\color{blue}$\textit{take} / f$}} (A);

  %\path[draw=brown,font=\tiny]      (C) edge [bend right] node[above,sloped] {{\color{brown}$\textit{cancel} / c$}} (D)
   %     (D) edge [bend right] node[above,sloped] {{\color{brown}$\textit{return} / c$}} (A);

\end{tikzpicture}
\vspace{-2.5mm}
\caption{$\pi_{\{\fv,\fs\}}(\small \textsc{VendingMachine})$}
\label{fig:variant1}
\end{minipage}
%\vspace{-2.5mm}
%\caption{The \vending\, variational system.}
%\label{fig:vending}
\end{figure}

\begin{example} \label{exp:1}
Throughout this paper, we will use a beverage vending machine as a running example \cite{model-checking-spls}.
Figure~\ref{fig:FTS} shows the FTS of a \vending \, family.
It has five features, and each of them is assigned an identifying letter and a color. The features are:
\texttt{VendingMachine} (denoted by letter $v$, in black), the mandatory base feature of purchasing a drink, present in all variants;
\texttt{Tea} (\ft, \textcolor{red}{in red}), for serving tea;
\texttt{Soda} (\fs, \textcolor{darkgreen}{in green}), for serving soda, which is a mandatory feature present in all variants;
\texttt{CancelPurchase} (\fc, \textcolor{brown}{in brown}), for canceling a purchase after a coin is entered; and
\texttt{FreeDrinks} (\ff, \textcolor{blue}{in blue}) for offering free drinks.
Each transition is labeled by an \emph{action} followed by a \emph{feature expression}. %specifying for which variants the transition is included.
For instance, the transition \ST 1 \textcolor{blue}{\TR{\textit{free}/\!f}} \ST 3 is included in variants where the feature \textcolor{blue}{$f$} is enabled.

By combining various features, a number of variants of this \vending\,
can be obtained.
Recall that \fv\, and \fs\, are mandatory features. The set of valid configurations is thus:
  \noindent$\displaystyle\Kk^{\textsc{VM}} \!=\! \{ \{ \fv,\linebreak[0] \fs \},\linebreak[0] \{ \fv,\linebreak[0] \fs,\linebreak[0] \ft \},\linebreak[0] \{\fv,\linebreak[0] \fs,\linebreak[0] \fc\},\linebreak[0] \{\fv,\linebreak[0] \fs,\linebreak[0] \ft,\linebreak[0] \fc\},\linebreak[0] \{ \fv,\linebreak[0] \fs,\linebreak[0] \ff \},\linebreak[0] \{ \fv,\linebreak[0] \fs,\linebreak[0] \ft,\linebreak[0] \ff \},\linebreak[0]  \{\fv,\linebreak[0] \fs,\linebreak[0] \fc,\linebreak[0] \ff \},\linebreak[0] \{\fv,\linebreak[0] \fs,\linebreak[0] \ft,\linebreak[0] \fc,\linebreak[0] \ff \} \}$.
In Fig.~\ref{fig:variant1} is shown the basic version of \vending\, that only serves soda, which is
described by the configuration: $\{v,s\}$ (or, as formula $v \land\! s \land\! \neg t \land\! \neg c \land\! \neg f$), that is
the projection \(\pi_{\{\fv,\fs\}}(\textsc{VendingMachine})\).
It takes a coin, returns change, serves soda, opens a compartment
so that the customer can take the soda, before closing it again.
%The model presented in Fig.~\ref{fig:variant1} is obtained by the projection \(\pi_{\{\fv,\fs\}}(\textsc{VendingMachine})\).

Figures~\ref{fig:abs1} and \ref{fig:abs4} show an MTS and an MFTS, respectively. Must transitions are denoted by solid lines,
may transitions by dashed lines. The MFTS in Fig.~\ref{fig:abs4} (Appendix~\ref{app:figures}) has $\Ff = \{\fc\}$ and $\Kk=\{\fc,\neg \fc\}$.
%For clarity, we omit to write the presence condition \true\, in transitions.
\qed
\end{example}

\paragraph{\textbf{CTL$^{\star}$ Properties.}}
Computation Tree Logic$^{\star}$ (CTL$^{\star}$)\,\cite{katoen-beier} is an expressive temporal logic for specifying system properties,
 which subsumes both CTL and LTL logics.
CTL$^{\star}$ state formulae $\Phi$ are generated by the following grammar:
\[
\Phi ::= \true \mid a \in \AP \mid \neg a \mid \Phi_1 \land \Phi_2 \mid  \forall \phi \mid \exists \phi, \qquad
\phi ::=  \Phi \mid \phi_1 \land \phi_2 \mid \bigcirc \phi \mid \phi_1 \U \phi_2
\]
where $\phi$ represent CTL$^{\star}$ path formulae.
Note that the CTL$^{\star}$ state formulae $\Phi$ are given in negation normal form
($\neg$ is applied only to atomic propositions).
Given $\Phi \in \textrm{CTL}^{\star}$, we consider $\neg \Phi$ to be the equivalent
CTL$^{\star}$ formula given in negation normal form.
Other derived temporal operators (path formulae) can be defined as well by means of syntactic sugar, for instance:
$\Diamond \phi = true \, \U \phi$ ($\phi$ holds eventually),
 and $\Box \phi = \neg \forall \Diamond \neg \phi$ ($\phi$ always holds).
$\forall$CTL$^{\star}$ and $\exists$CTL$^{\star}$ are subsets of CTL$^{\star}$ where the
only allowed path quantifiers are $\forall$ and $\exists$, respectively.

 We formalise the semantics of CTL$^{\star}$ over a TS $\Ttt$.  We write $\sbr{\Ttt}^{s}_{\textrm{TS}}$ for the set of executions that start in state $s$; $\rho[i]=s_i$ to denote the $i$-th state of the execution $\rho$; and
 $\rho_i = s_i \lambda_{i+1} s_{i+1} \ldots$ for the suffix of $\rho$ starting from its $i$-th state.
\begin{definition} \label{def:ctl}
 Satisfaction of a state formula $\Phi$ in a state $s$ of a TS $\Ttt$, denoted $\Ttt,s \models \phi$, is defined as ($\Ttt$
is omitted when clear from context):
\begin{description}
\item[(1)] $s \models a$ iff $a \in L(s)$; $s \models \neg a$ iff $a \notin L(s)$,
\item[(2)] $s \models \Phi_1 \land \Phi_2$ iff $s \models \Phi_1$ and $s \models \Phi_2$,
\item[(3)] $s \models \forall \phi$ iff  $\forall \rho \in \sbr{\Ttt}^{s}_{\textrm{TS}}. \, \rho \models \phi$;
$s \models \exists \phi$ iff  $\exists \rho \in \sbr{\Ttt}^{s}_{\textrm{TS}}. \, \rho \models \phi$
\end{description}
Satisfaction of a path formula $\phi$ for an execution $\rho$ of a TS $\Ttt$, denoted $\Ttt,\rho \models \phi$, is
defined as ($\Ttt$ is omitted when clear from context):
\begin{description}
\item[(4)] $\rho \models \Phi$ iff $\rho[0] \models \Phi$,
\item[(5)] $\rho \models \phi_1 \land \phi_2$ iff $\rho \models \phi_1$ and $\rho \models \phi_2$;
$\rho \models \bigcirc \phi$ iff $\rho_1 \models \phi$;
$\rho \models (\phi_1 \U \phi_2)$ iff $\exists i \!\geq\! 0. \, \big( \rho_i \models \phi_2 \land (\forall 0 \!\leq\! j \!\leq\! i\!-\!1. \, \rho_j \models \phi_1) \big)$
\end{description}
A TS $\Ttt$ satisfies a state formula $\Phi$, written $\Ttt \models \Phi$, iff %all its initial states satisfy the formula:
$\forall s_0 \in I. \, s_0 \models \Phi$.
\end{definition}

\begin{definition}
An FTS $\Fff$ satisfies a CTL$^{\star}$ formula $\Phi$, written $\Fff \models \Phi$, iff all its valid variants satisfy the formula:
$  \forall k\!\in\!\Kk. \, \pi_k(\Fff) \models \Phi$.
\end{definition}

The interpretation of CTL$^{\star}$ over an MTS $\Mmm$ is defined slightly different from the above Definition~\ref{def:ctl}. In particular,
the clause (3) is replaced by:
\begin{description}
\item[(3')] $s \models \forall \phi$ iff for every may-execution $\rho$ in the state $s$ of $\Mmm$, that is $\forall \rho \in \sbr{\Mmm}_{MTS}^{may,s}$, it holds $\rho \models \phi$; whereas
$s \models \exists \phi$ iff there exists a must-execution $\rho$ in the state $s$ of $\Mmm$, that is $\exists \rho \in \sbr{\Mmm}_{MTS}^{must,s}$, such that $\rho \models \phi$.
\end{description}
From now on, we implicitly assume this adapted definition when interpreting CTL$^{\star}$ formulae over MTSs and MFTSs.

\begin{example} \label{exp:property}
Consider the FTS \vending\, in Fig.~\ref{fig:FTS}.
Suppose that the proposition $\texttt{start}$ holds in the initial state $\ST 1$.
An example property $\Phi_1$ is: $\forall \Box \, \forall \Diamond \textnormal{\texttt{start}}$, which states that in every state
along every execution all possible continuations will eventually reach the initial state. This formula is in $\forall$CTL$^{\star}$.
Note that $\vending \not\models \Phi_1$.
For example, if the feature $\fc$ (\texttt{Cancel})  is enabled, a counter-example where the state $\ST 1$ is never reached is:
$\ST 1 \to \ST 3 \to \ST 5 \to \ST 7 \to \ST 3 \to \ldots$. The set of violating products is
$\sbr{c}\!=\!\{ \{ \fv,\linebreak[0] \fs,\linebreak[0] \fc \},\linebreak[0] \{ \fv,\linebreak[0] \fs,\linebreak[0] \ft,\linebreak[0] \fc \},\linebreak[0]  \{\fv,\linebreak[0] \fs,\linebreak[0] \fc,\linebreak[0] \ff \},\linebreak[0] \{\fv,\linebreak[0] \fs,\linebreak[0] \ft,\linebreak[0] \fc,\linebreak[0] \ff \} \} \subseteq \Kk^{\textsc{VM}}$.
However, $\pi_{\sbr{\neg \fc}}(\vending) \models \Phi_1$.

Consider the property $\Phi_2$: $\forall \Box \, \exists \Diamond \textnormal{\texttt{start}}$, which describes a situation where
in every state
along every execution there exists a possible continuation that will eventually reach the \texttt{start} state.
This is a CTL$^{\star}$ formula, which is neither in $\forall$CTL$^{\star}$ nor in $\exists$CTL$^{\star}$.
Note that $\vending \models \Phi_2$, since even for variants with the feature $\fc$ there is a continuation from the state
$\ST 3$ back to $\ST 1$.

Consider the $\exists$CTL$^{\star}$ property $\Phi_3$: $\exists \Box \, \exists \Diamond \textnormal{\texttt{start}}$, which states that
there exists an execution such that in every state
along it there exists a possible continuation that will eventually reach the \texttt{start} state.
The witnesses are
%$\ST 1 \to \ST 3 \to \ST 5 \to \ST 7 \to \ST 8 \to \ST 1 \ldots$ and
$\ST 1 \to \ST 2 \to \ST 3 \to \ST 5 \to \ST 7 \to \ST 8 \to \ST 1 \ldots$ for variants that satisfy $\neg \fc$,
and $\ST 1 \to \ST 3 \to \ST 5 \to \ST 7 \to \ST 3 \to \ST 4 \to \ST 1 \ldots$ for variants with $\fc$.
\qed
\end{example}

\section{Abstraction of FTSs}\label{sec:definition}

We now introduce the variability abstractions which
preserve full CTL and its universal and existential properties.
They simplify the configuration space of an FTSs, by reducing the number of configurations and
 manipulating presence conditions of transitions.
We start working with Galois connections \footnote{$\poset{L}{\leq_L} \galois{\alpha}{\gamma} \poset{M}{\leq_M}$ is a \emph{Galois connection} between complete lattices $L$ (concrete domain) and $M$ (abstract domain) iff $\alpha:L\to M$ and $\gamma:M \to L$ are total functions that satisfy: $\alpha(l) \leq_M m \iff l \leq_L \gamma(m)$ for all $l \in L, m \in M$.
Here $\leqslant_L$ and $\leqslant_M$ are the pre-order relations
for $L$ and $M$, respectively. We will often simply write $(\alpha,\gamma)$ for any such Galois connection.
 } between Boolean complete lattices of feature expressions, and then induce a notion of abstraction of FTSs.
We define two classes of abstractions. We use the standard conservative abstractions \cite{spin15,sttt16} as an instrument to eliminate variability from the FTS in an \emph{over-approximating} way, so by adding more executions.  We use the dual abstractions, which can also eliminate variability but through \emph{under-approximating} the given FTS, so by dropping executions. %We now present these two classes of abstractions.

\paragraph{\textbf{Domains.}}
The Boolean complete lattice of feature expressions (propositional formulae over $\Ff$) is: \((\FeatExp(\Ff)_{/\equiv},\models,\lor,\land,\true,\false,\neg)\).  The elements of the domain
$\FeatExp(\Ff)_{/\equiv}$ are equivalence classes of propositional formulae \(\psi\!\in\! \FeatExp(\Ff)\) obtained by quotienting by the semantic equivalence
$\equiv$. The ordering $\models$ is the standard entailment between propositional logics formulae,
% Note: it is not satisfaction; satisfaction is between model and a formula; entailment is between formulae.
whereas the least upper bound and the greatest lower bound are just logical disjunction and conjunction respectively.
Finally, the constant \false\ is the least, \true\ is the greatest element, and negation is the complement operator.

\paragraph{\textbf{Conservative abstractions.}}
The \emph{join abstraction}, $\joinasym$, merges the control-flow of all variants, obtaining a single variant that includes all executions occurring in any variant.  The information about which transitions are associated with which variants is lost.  Each feature expression $\psi$ is replaced with \true\ if there exists at least one configuration from $\Kk$ that satisfies $\psi$.
The new abstract set of features is empty: $\joinasym(\Ff)=\emptyset$, and the abstract set of valid configurations is a singleton: $\joinasym(\Kk) = \{ \true \}$ if $\Kk \neq \emptyset$.  %For the empty set of configurations, the abstraction is not defined.
The abstraction and concretization functions
between $\FeatExp(\Ff)$ and $\FeatExp(\emptyset)$, forming a Galois connection \cite{spin15,sttt16}, are defined as:
\[
\joinasym(\psi) \!=\!
  \begin{cases}
    \true  & \textrm{if } \exists k \in \Kk. k \models \psi \\
    \false & \textrm{otherwise}
  \end{cases} \qquad \
\joingsym(\psi) \!=\!
  \begin{cases}
  \true & \text{if } \psi \text{ is } \true\\
  \bigvee_{k \in 2^\Ff\!\setminus\! \Kk} k & \text{if } \psi \text{ is } \false
  \end{cases}
\]

The \emph{feature ignore abstraction}, $\fignorea{A}$, %is a feature-oriented counterpart for the join abstraction.  It
introduces an over-approximation by ignoring a single feature $A\!\in\!\Ff$. It  merges the control flow paths that only differ with regard to $A$, but keeps the precision with respect to control flow paths that do not depend on $A$. %The feature $A$ is entirely eliminated from the system.
The features and configurations of the abstracted model are:
$\fignorea{A}(\Ff) = \Ff \!\setminus\!\{A\}$, and
$\fignorea{A}(\Kk) = \{ k[l_A \mapsto \true] \mid k\in\Kk \}$,
where $l_A$ denotes a literal of $A$ (either $A$ or $\neg A$), and $k[l_A\mapsto\true]$ is a formula resulting from $k$ by substituting \true\ for $l_A$.  %We assume that the configurations $k$ are written in the negation normal form. % (which can always be achieved without changing the semantics of the formula).
The abstraction and concretization functions between $\FeatExp(\Ff)$ and $\FeatExp(\fignorea{A}(\Ff))$, forming a Galois connection \cite{spin15,sttt16}, are:
\[
  \fignorea{A}(\psi) = \psi[l_A \mapsto \true] \qquad \
  \fignoreg{A}(\psi') = (\psi' \land A) \lor (\psi' \land \neg A)
\]
where $\psi$ and $\psi'$ need to be in negation normal form before substitution.
%The feature ignore abstraction is directly generalizable to sets of features \cite{spin15,sttt16}.
%Note that, $\joinasym$ can be obtained as an abstraction that ignores all available features,
%that is $\joinasym = \fignorea{\Ff}$.
%In the following, we will often simply write $(\alpha,\gamma)$ for any Galois connection $\poset{\FeatExp(\Ff)_{/\equiv}}{\models} \galois{\alpha}{\gamma} \poset{\FeatExp(\alpha(\Ff))_{/\equiv} }{\models}$ to conserve space.

\paragraph{\textbf{Dual abstractions.}}
Suppose that $\poset{\FeatExp(\Ff)_{/\equiv}}{\models}$, $\poset{\FeatExp(\alpha(\Ff))_{/\equiv} }{\models}$ are Boolean complete lattices, and  $\poset{\FeatExp(\Ff)_{/\equiv}}{\models} \galois{\alpha}{\gamma} \poset{\FeatExp(\alpha(\Ff))_{/\equiv} }{\models}$ is a Galois connection.
We define \cite{DBLP:conf/sara/Cousot00}:
$\widetilde{\alpha}=\neg \circ \alpha \circ \neg$ and $\widetilde{\gamma}=\neg \circ \gamma \circ \neg$ so that
$\poset{\FeatExp(\Ff)_{/\equiv}}{\ndststile{}{}} \galois{\widetilde{\alpha}}{\widetilde{\gamma}} \poset{\FeatExp(\alpha(\Ff))_{/\equiv} }{\ndststile{}{}}$ is a Galois connection
(or equivalently, $\poset{\FeatExp(\alpha(\Ff))_{/\equiv}}{\models} \galois{\widetilde{\gamma}}{\widetilde{\alpha}} \poset{\FeatExp(\Ff)_{/\equiv} }{\models}$).
The obtained Galois connections $(\widetilde{\alpha},\widetilde{\gamma})$ are called dual (under-approximating) abstractions of $(\alpha,\gamma)$.

The \emph{dual join abstraction}, $\widetilde{\joinasym}$, merges the control-flow of all variants, obtaining a single variant that includes only those executions that occur in all variants. Each feature expression $\psi$ is replaced with \true\ if all configurations from $\Kk$ satisfy $\psi$.
The abstraction and concretization functions
between $\FeatExp(\Ff)$ and $\FeatExp(\emptyset)$, forming a Galois connection, are defined as:
$\widetilde{\joinasym}=\neg \circ \joinasym \circ \neg$ and $\widetilde{\joingsym}=\neg \circ \joingsym \circ \neg$, that is:
\begin{equation*}
\widetilde{\joinasym}(\psi) =
\begin{cases}
    \true & \textrm{if } \forall k \in \Kk. k \models \psi \\
	\false  & \textrm{otherwise }
	
\end{cases}
\qquad
\widetilde{\joingsym}(\psi) \!=\!
  \begin{cases}
  \bigwedge_{k \in 2^\Ff \backslash \Kk} (\neg k) & \text{if } \psi \text{ is } \true\\
  \false & \text{if } \psi \text{ is } \false
  \end{cases}
\end{equation*}

The \emph{dual feature ignore abstraction}, $\widetilde{\fignorea{A}}$,
introduces an under-approximation by ignoring the feature $A\!\in\!\Ff$, such that the literals of $A$ (that is, $A$ and $\neg A$)
are replaced with \false\, in feature expressions (given in negation normal form).
The abstraction and concretization functions between $\FeatExp(\Ff)$ and $\FeatExp(\fignorea{A}(\Ff))$, forming a Galois connection, are defined as:
$\widetilde{\fignorea{A}}=\neg \circ \fignorea{A} \circ \neg$ and $\widetilde{\fignoreg{A}}=\neg \circ \fignoreg{A} \circ \neg$, that is:
\[
\widetilde{\fignorea{A}}(\psi) = \psi[l_A \mapsto \false] \qquad \widetilde{\fignoreg{A}}(\psi') = (\psi' \lor \neg A) \land (\psi' \lor A)
\]
where $\psi$ and $\psi'$ are in negation normal form.

\paragraph{\textbf{Abstract MFTS and Preservation of CTL$^{\star}$.}}
Given a Galois connection $(\alpha,\gamma)$ defined on the level of feature expressions,
we now define the abstraction of an FTS as an MFTS with two transition relations: one (may)
preserving universal properties, and the other (must) existential properties.
The may transitions describe the behaviour that is possible, but not need be realized in the
variants of the family; whereas the must transitions describe behaviour that has to be present
in any variant of the family.
\begin{definition}
Given the FTS $\Fff=(S,Act,trans,I,AP,L,\Ff,\Kk,\delta)$, %$[\chi] \phi$ be an fLTL formula,
%and $(\alpha,\gamma)$ be a Galois connection.
we define the MFTS $\alpha(\Fff)=(S,Act,trans^{may},trans^{must},I,AP,L,\alpha(\Ff),\alpha(\Kk),\delta^{may},\delta^{must})$
to be its abstraction,
where  $\delta^{may}(t)=\alpha(\delta(t))$, $\delta^{must}(t)=\widetilde{\alpha}(\delta(t))$,
 $trans^{may} = \{ t \in trans \mid \delta^{may}(t) \neq \false \}$, and
 $trans^{must} = \{ t \in trans \mid \delta^{must}(t) \neq \false \}$.
%\item We define $\alpha([\chi] \phi) = [\alpha(\chi)] \phi$.
\end{definition}
Note that the degree of reduction is determined by the choice of abstraction and may hence be arbitrary large.
In the extreme case of join abstraction, we obtain an abstract model with no variability in it,
that is $\joinasym(\Fff)$ is an ordinary MTS.

\begin{example}
Recall the FTS $\vending$ of Fig.~\ref{fig:FTS} with the set of valid configurations $\Kk^{\textsc{VM}}$ (see Example \ref{exp:1}).
Fig.~\ref{fig:abs1} shows $\joinasym({\vending})$, where the allowed (may) part of the behavior includes  the transitions that are associated with the optional features \fc, \ff, \ft\, in \vending, whereas the required (must) part includes the transitions associated with the mandatory features \fv\, and \fs.
Note that $\joinasym({\vending})$ is an ordinary MTS with no variability.
The MFTS $\fignorea{\{\ft,\ff\}}(\pi_{\sbr{\fv \,\land\, \fs}}(\textsc{VendingMachine}))$ is shown in Fig.~\ref{fig:abs4} (Appendix~\ref{app:figures}).
It has the singleton set of features $\Ff=\{\fc\}$ and limited variability $\Kk=\{\fc,\neg \fc\}$, where the mandatory features \fv\, and \fs\,
  are enabled. \qed
\end{example}

From the MFTS (resp., MTS) $\Mff$, we define two FTSs (resp., TSs) $\Mff^{may}$ and $\Mff^{must}$ representing the may- and must-components
of $\Mff$, i.e.\ its may and must transitions, respectively.
Thus, we have $\sbr{\Mff^{may}}_{FTS}=\sbr{\Mff}^{may}_{MFTS}$ and $\sbr{\Mff^{must}}_{FTS}=\sbr{\Mff}^{must}_{MFTS}$.

We now show that the abstraction of an FTS is sound with respect to CTL$^{\star}$.
First, we show two helper lemmas stating
that: for any variant $k \!\in\! \Kk$ that can execute a behavior, there exists an abstract variant $k' \!\in\! \alpha(\Kk)$
that executes the same may-behaviour; and for any abstract variant $k' \!\in\! \alpha(\Kk)$ that can execute a must-behavior, there exists a variant $k \!\in\! \Kk$ that executes the same behaviour \footnote{Proofs of all lemmas and theorems in this section can be found in Appendix~\ref{app:proofs}.}.

\begin{lemma} \label{lemma:1}
Let $\psi  \in \FeatExp(\Ff)$, and $\Kk$ be a set of valid configurations over $\Ff$.
\begin{description}
\item[(i)] Let $k \in \Kk$ and $k \models \psi$. Then there exists $k' \in \alpha(\Kk)$, such that $k' \models \alpha(\psi)$.
\item[(ii)] Let $k' \in \alpha(\Kk)$ and $k' \models \widetilde{\alpha}(\psi)$. Then for all $k \in \Kk$ s.t. $\alpha(k)=k'$, it holds $k \models \psi$.
\end{description}
\end{lemma}
%\begin{proof}
%By induction on the structure of $\alpha$. See Appendix~\ref{app:proofs}.
%\end{proof}

\begin{lemma} \label{lemma:2}
\begin{description}
\item[(i)] Let $k \in \Kk$ and $\rho \in \sbr{\pi_k(\Fff)}_{TS} \subseteq \sbr{\Fff}_{FTS}$. Then there exists $k' \in \alpha(\Kk)$, such that $\rho \in \sbr{\pi_{k'}(\alpha(\Fff))}^{may}_{MTS} \subseteq \sbr{\alpha(\Fff)}^{may}_{MFTS}$ is a may-execution in $\alpha(\Fff)$.
\item[(ii)] Let $k' \in \alpha(\Kk)$ and $\rho \in \sbr{\pi_{k'}(\alpha(\Fff))}^{must}_{MTS} \subseteq \sbr{\alpha(\Fff)}^{must}_{MFTS}$ be a must-execution in $\alpha(\Fff)$. Then for all $k \in \Kk$ s.t. $\alpha(k)=k'$, it holds $\rho \in \sbr{\pi_k(\Fff)}_{TS} \subseteq \sbr{\Fff}_{FTS}$.
\end{description}
\end{lemma}
%\begin{proof}
%By definition of FTSs, MFTSs, and Lemma~\ref{lemma:1}. See Appendix~\ref{app:proofs}.
%\end{proof}

As a result, every $\forall$CTL$^{\star}$ (resp., $\exists$CTL$^{\star}$) property true for the may-
(resp., must-) component of $\alpha(\Fff)$ is true for $\Fff$ as well. Moreover, the MFTS $\alpha(\Fff)$
preserves the full CTL$^{\star}$.

\begin{theorem}[Preservation results] \label{theorem:sound}
For any FTS $\Fff$ and $(\alpha,\gamma)$, we have:
\begin{description}
\item[($\forall$CTL$^{\star}$)] For every $\Phi \in \forall CTL^{\star}$,
$\alpha(\Fff)^{may} \models \Phi \ \implies \ \Fff \models \Phi$.
\item[($\exists$CTL$^{\star}$)] For every $\Phi \in \exists CTL^{\star}$,
$\alpha(\Fff)^{must} \models \Phi \ \implies \ \Fff \models \Phi$.
\item[(CTL$^{\star}$)]
For every $\Phi \in CTL^{\star}$,
$\alpha(\Fff) \models \Phi \ \implies \ \Fff \models \Phi$.
\end{description}
\end{theorem}
%\begin{proof}
%By induction on the structure of $\Phi$ and Lemma~\ref{lemma:2}. See Appendix~\ref{app:proofs}.
%\end{proof}

Abstract models are designed to be conservative for the satisfaction of properties.
However, in case of the refutation of a property, a counter-example is found in the abstract model which
may be spurious (introduced due to abstraction) for some variants and genuine for the others.
This can be established by checking which variants can execute the found counter-example.

\begin{figure}[t]
\centering
\begin{tikzpicture}[->,>=stealth',shorten >=0.9pt,auto,node distance=1.45cm, semithick]
  \tikzstyle{every state}=[minimum size=.2pt,initial text={{\scriptsize start}}]

  \node[initial below,state] (A)     {\tiny 1};
  \node[state]         (B) [right of=A] {\tiny 2};
  \node[state]         (C) [right of=B] {\tiny 3};
  \node[state]         (D) [above right of=A]  {\tiny 4};
  \node[state]         (E) [above right of=C]  {\tiny 5};
  \node[state]         (F) [below right=4mm of C]  {\tiny 6};
  \node[state]         (G) [below right of=E]  {\tiny 7};
  \node[state]         (H) [right of=G]  {\tiny 8};
  %\node[state]         (I) [right of=H]  {\tiny 9};

  \path[font=\tiny] (A) edge              node {$\textit{pay}$} (B)
        (B) edge              node {$\textit{change}$} (C)
        (G) edge   node {$\textit{open}$} (H)
        %(H) edge   node [above,sloped]  {$\textit{take}$} (I)
        (H) edge [in=-70,out=-25]  node[above right,sloped] {$\textit{take}$} (A);

  \path[font=\tiny]      (C) edge [bend left] node[above,sloped] {$\textit{soda}$} (E)
        (E) edge [bend left] node[above,sloped] {$s\textit{erveSoda}$} (G);

  \path[draw, dashed, font=\tiny]      (C) edge [bend right] node[below,sloped] {$\textit{tea}$} (F)
        (F) edge [bend right, in=-180] node[above,sloped] {$\textit{~~serveTea}$} (G);

  \path[draw, dashed, font=\tiny]      (A) edge [bend right] node[below,sloped] {$\textit{free}$} (C)
        (G) edge [bend left=56,in=128,out=128] node[above,sloped] {$\textit{take}$} (A)
        (G) edge [in=68,out=128] node[below,sloped] {$\textit{take}$} (C);

  \path[draw, dashed, font=\tiny]      (C) edge [bend right] node[above,sloped] {$\textit{cancel}$} (D)
        (D) edge [bend right] node[above,sloped] {$\textit{return}$} (A);

\end{tikzpicture}
\vspace{-11mm}
\caption{$\joinasym(\textsc{VendingMachine})$.}
\label{fig:abs1}
\end{figure}

Let $\Phi$ be a CTL$^{\star}$ formula which is not in $\forall$CTL$^{\star}$ nor in $\exists$CTL$^{\star}$, and
let $\Mff$ be an MFTS. We verify $\Mff \models \Phi$ by checking $\Phi$ on two FTSs $\Mff^{may}$ and $\Mff^{must}$,
and then we combine the obtained results as specified below.

\begin{theorem}\label{theorem:mfts}
For every  $\Phi \in CTL^{\star}$ and MFTS $\Mff$, we have:
\[
\Mff \models \Phi = \begin{cases} \true & \textrm{if  } \big( \Mff^{may} \models \Phi \, \land \, \Mff^{must} \models \Phi \big) \\
\false & \textrm{if  } \big( \Mff^{may} \not\models \Phi \, \lor \, \Mff^{must} \not\models \Phi \big) \end{cases}
\]
\end{theorem}
Therefore, we can check a %CTL$^{\star}$
formula $\Phi$ which is not in $\forall$CTL$^{\star}$ nor in $\exists$CTL$^{\star}$ on $\alpha(\Fff)$ by running a model checker twice, once with the may-component of $\alpha(\Fff)$ and once with the must-component of $\alpha(\Fff)$.
On the other hand, a formula $\Phi$ from $\forall$CTL$^{\star}$ (resp., $\exists$CTL$^{\star}$) on
 $\alpha(\Fff)$ is checked by running a model checker only once with the may-component (resp., must-component) of $\alpha(\Fff)$.

The family-based model
checking problem can be reduced to a number of smaller problems by
partitioning the set of variants.
Let the subsets $\Kk_1, \Kk_2, \ldots, \Kk_n$ form a \emph{partition} of the set $\Kk$.
Then: $\Fff \models \Phi$ iff $\pi_{\Kk_i}(\Fff) \models \Phi$ for all $i=1,\ldots,n$.
By using Theorem~\ref{theorem:sound} (CTL$^\star$), we obtain the following result.
\begin{corollary}
Let $\Kk_1, \Kk_2, \ldots, \Kk_n$ form a \emph{partition} of $\Kk$,
and $(\alpha_1,\!\gamma_1), \ldots, (\alpha_n,\!\gamma_n)$ be Galois connections.
If $\alpha_1(\pi_{\Kk_1}(\Fff)) \models \Phi, \ldots, \alpha_n(\pi_{\Kk_n}(\Fff)) \models \Phi$,
 then $\Fff \models \Phi$.
\end{corollary}
Therefore, in case of suitable partitioning of $\Kk$ and the aggressive $\joinasym$ abstraction, all  $\joinasym({\pi_{\Kk_i}(\Fff)})^{may}$ and $\joinasym({\pi_{\Kk_i}(\Fff)})^{must}$ are
ordinary TSs, so the family-based model checking problem can be solved using existing single-system model checkers with all
the optimizations that these tools may already implement.

\begin{example}
Consider the properties introduced in Example~\ref{exp:property}.
Using the TS $\joinasym({\vending})^{may}$ we can verify $\Phi_1=\forall \Box \, \forall \Diamond \textnormal{\texttt{start}}$ (Theorem~\ref{theorem:sound}, ($\forall$CTL$^{\star}$)).
We obtain the counter-example $\ST 1 \to \ST 3 \to \ST 5 \to \ST 7 \to \ST 3 \ldots$, which is genuine for variants satisfying $\fc$.
Hence, variants from $\sbr{\fc}$ violate $\Phi_1$.
On the other hand,
by verifying that $\joinasym({\pi_{\sbr{\neg \fc}}(\vending)})^{may}$ satisfies $\Phi_1$, we can conclude by
Theorem~\ref{theorem:sound}, ($\forall$CTL$^{\star}$) that variants from $\sbr{\neg \fc}$
satisfy $\Phi_1$.

We can verify $\Phi_2=\forall \Box \, \exists \Diamond \textnormal{\texttt{start}}$ by checking may- and must-components
of $\joinasym(\vending)$. In particular, we have $\joinasym(\vending)^{may} \models \Phi_2$ and
$\joinasym(\vending)^{must} \models \Phi_2$. Thus, using Theorem~\ref{theorem:sound}, (CTL$^{\star}$) and
Theorem~\ref{theorem:mfts}, we have that $\vending \models \Phi_2$.

Using $\joinasym({\vending})^{must}$ we can verify $\Phi_3=\exists \Box \, \exists \Diamond \textnormal{\texttt{start}}$, by
finding the witness $\ST 1 \to \ST 2 \to \ST 3 \to \ST 5 \to \ST 7 \to \ST 8 \to \ST 1 \ldots$.
By Theorem~\ref{theorem:sound}, ($\exists$CTL$^{\star}$), we have that $\vending \models \Phi_3$. \qed
\end{example} 

\section{Implementation}
%\section{Abstracting fPromela+TVL}

We now describe an implementation of our abstraction-based approach for
CTL model checking of variational systems in the context of the state-of-the-art \smv\,
model checker \cite{DBLP:conf/cav/CimattiCGGPRST02}.
Since it is difficult to use FTSs to directly model very large variational
systems, we use a high-level modelling language, called \fsmv,
which is expressively equivalent to FTSs and close to \smv's input language.
Then, we show how to implement projection and variability abstractions as
syntactic transformations of \fsmv\, models.

\paragraph{\textbf{A High-level Modelling Language.}}
\fsmv\, is a feature-oriented extension of the input language of \smv, which was
introduced by Plath and Ryan \cite{DBLP:journals/scp/PlathR01} and subsequently improved by Classen \cite{classenTR}.
A \smv\, model consists of
a set of variable declarations and a set of assignments.
The variable declarations define the state space and the assignments define
the transition relation of the finite state machine described by the given model.
For each variable, there are assignments that define its initial value and its value in the
next state, which is given as a function of the variable values in the present state.
Modules can be used to encapsulate and factor out recurring submodels.
Consider a basic \smv\, model shown in Fig.~\ref{fig:fsmv1}.
It consists of a single variable $x$ which is initialized to 0 and does not change its value.
The property (marked by the keyword \texttt{SPEC}) is ``$\forall \Diamond (x \geq k)$'', where $k$ is a meta-variable that can be
replaced with various natural numbers.
For this model, the property holds when $k=0$.
In all other cases (for $k > 0$), %the property is violated and
a counterexample is reported where $x$ stays 0.

The \fsmv\, language \cite{DBLP:journals/scp/PlathR01} is based on superimposition.
\emph{Features} are modelled as self-contained textual units using a new \texttt{FEATURE} construct
added to the \smv\, language.
A feature describes the changes to be made
to the given basic \smv\, model. It can introduce new variables into the system (in a section marked by the keyword \texttt{INTRODUCE}),
override the definition of existing variables in the basic model and change the values of those variables
when they are read (in a section marked by the keyword \texttt{CHANGE}).
For example, Fig.~\ref{fig:fsmv3} shows a \texttt{FEATURE} construct, called $A$,
which changes the basic model in Fig.~\ref{fig:fsmv1}.
In particular, the feature $A$ defines a new variable $nA$ initialized to 0.
The basic system is changed in such a way that when
the condition ``$nA=0$'' holds then in the next state the basic system's variable $x$ is
incremented by 1 and in this case (when $x$ is incremented) $nA$ is set to 1.
Otherwise, the basic system is not changed.

Classen \cite{classenTR} shows
that \fsmv\, and FTS are expressively equivalent.
He \cite{classenTR} also proposes a way of composing \fsmv\, features with the basic model to create
a single model in pure \smv\, which describes all valid variants.
The information about the variability and features in the composed
model is recorded in the states. This is a slight deviation from the encoding
in FTSs, where this information is part of the transition relation. However, this encoding
has the advantage of being implementable in \smv\, without drastic changes
 to the model checker and its input language.
Basically,
in the composed model each feature becomes a Boolean state variable,
which is non-deterministically initialised and whose value never changes. % by the transitions.
Thus, the initial states of the composed model include all possible feature combinations.
Every change
performed by a feature in the composition
is guarded by the corresponding feature variable.

For example, the composition of the basic model and the feature $A$ given in Figs.~\ref{fig:fsmv1} and \ref{fig:fsmv3}
results in the model shown in  Fig.~\ref{fig:fsmv2}.
First, a module, called \texttt{\emph{features}}, containing all features (in this case, the single one $A$) is added to the system.
To each feature (e.g. $A$) corresponds one variable in this module (e.g. $fA$).
The \texttt{\emph{main}} module contains a variable named $f$ of type \texttt{\emph{features}}, so that all feature
variables can be referenced in it (e.g. $f.fA$).
In the next state, the variable $x$ is incremented by 1 when the feature $A$ is enabled ($fA$ is \textit{TRUE})
and $nA$ is 0. Otherwise (\textit{TRUE:} can be read as \textit{else:}), $x$ is not changed.
Also, $nA$ is set to 1 when $A$ is enabled and $x$ is incremented by 1.
The property $\forall \Diamond (x \geq 0)$ holds for both variants when $A$ is enabled and $A$ is disabled ($fA$ is \textit{FALSE}).

\begin{figure}[t]
\begin{minipage}[b]{.4\textwidth}
\centering
\begin{subfigure}[b]{\linewidth}
$ \begin{array}{l}
\textcolor{gray}{\phantom{1}1}~~ \texttt{MODULE \emph{main} }\\
\textcolor{gray}{\phantom{1}2}~~ \texttt{VAR} \ x : 0..1;\\
\textcolor{gray}{\phantom{1}3}~~ \texttt{ASSIGN}\\
\textcolor{gray}{\phantom{1}4}~~ \quad \texttt{init}(x) := 0;\\
\textcolor{gray}{\phantom{1}5}~~ \quad \texttt{next}(x) := x; \\
\textcolor{gray}{\phantom{1}6}~~ \texttt{SPEC} \ AF (x \geq k);
\end{array} $
\vspace{-1mm}
\caption{The basic model.}
\label{fig:fsmv1}
\end{subfigure}\\[\baselineskip]
\begin{subfigure}[b]{\linewidth}
$\begin{array}{l}
\textcolor{gray}{\phantom{1}1}~~ \texttt{FEATURE} \, A \\
\textcolor{gray}{\phantom{1}2}~~ \texttt{INTRODUCE} \,  \\
\textcolor{gray}{\phantom{1}3}~~ \quad \texttt{VAR } nA : 0..1;\\
\textcolor{gray}{\phantom{1}4}~~ \quad \texttt{ASSIGN init}(nA) := 0;\\
\textcolor{gray}{\phantom{1}5}~~ \texttt{CHANGE} \, \\
\textcolor{gray}{\phantom{1}6}~~ \quad \texttt{IF} \, (nA=0) \, \texttt{THEN} \\
\textcolor{gray}{\phantom{1}7}~~ \quad \texttt{IMPOSE next}(x):=x+1; \\
\textcolor{gray}{\phantom{1}8}~~ \qquad \qquad \ \, \texttt{next}(nA):= \\
\textcolor{gray}{\phantom{1}9}~~ \qquad \quad {\small \texttt{next}(x)\!=\!x\!+\!1?1\!:\!nA;}
\end{array}$
\vspace{-1mm}
\caption{The feature A.}
\label{fig:fsmv3}
\end{subfigure}
\end{minipage}
\hfill
\begin{subfigure}[b]{.6\linewidth}
\centering
$ \begin{array}{l}
\textcolor{gray}{1}~~ \texttt{MODULE \emph{features} }\\
\textcolor{gray}{2}~~ \quad \texttt{VAR} \  fA : boolean;\\
\textcolor{gray}{3}~~ \quad \texttt{ASSIGN}\\
\textcolor{gray}{4}~~ \qquad \texttt{init}(fA) := \{\textit{TRUE,FALSE}\}; \\
\textcolor{gray}{5}~~ \qquad \texttt{next}(fA) := fA; \\
\textcolor{gray}{6}~~ \texttt{MODULE \emph{main} } \\
\textcolor{gray}{7}~~ \quad \texttt{VAR}\ f:features; \, x:0..1; \, nA : 0..1;\\
\textcolor{gray}{8}~~ \quad \texttt{ASSIGN}\\
\textcolor{gray}{9}~~ \qquad \ \texttt{init}(x) := 0; \texttt{init}(nA) := 0; \\
\textcolor{gray}{10}~~ \qquad \texttt{next}(x) := \texttt{case} \, f.fA \, \& \, nA\!=\!0 : x\!+\!1; \\
\textcolor{gray}{11}~~ \qquad \qquad \qquad \qquad \quad \textit{TRUE : } x; \\
\textcolor{gray}{12}~~ \qquad \qquad \qquad \quad \texttt{easc}; \\
\textcolor{gray}{13}~~ \qquad \texttt{next}(nA) := \texttt{case} \\
\textcolor{gray}{14}~~ \qquad \qquad f.fA \, \& \, nA\!=\!0 \, \& \, \texttt{next}(x)\!=\!x\!+\!1: 1; \\
\textcolor{gray}{15}~~ \qquad \qquad \textit{TRUE : } nA; \\
\textcolor{gray}{16}~~ \qquad \qquad \qquad \qquad \texttt{easc};
\end{array} $
\vspace{-1mm}
\caption{The composed model $\Mmm$.}
\label{fig:fsmv2}
\end{subfigure}
\vspace{-5.5mm}
\caption{\smv\ models.} \label{fig:fsmv}
\end{figure}

\begin{comment}
Suppose that there is another feature $fB$ which identically to $fA$ increases $i$ by 1.
If we use family-based model checking \cite{classen-model-checking-spls-icse11} to check
the property  $AF (i \geq 0)$ in the composed system, we will obtain that it is satisfied by
all variants. However, the property $AF (i > 0)$ will be violated by the variant $\neg fA \!\land\! \neg fB$,
where the value of $i$ stays 0 forever.
\end{comment}

\paragraph{\textbf{Transformations.}}
We present the syntactic transformations of \fsmv\ models defined by projection and variability abstractions.
Let $M$ represent a model obtained by composing a basic model with a set of features $\Ff$.
Let $M$ contain a set of assignments of the form:
$s(v) := \texttt{case } b_1:e_1; \ldots b_n:e_n; \texttt{ esac}$,
where $v$ is  a variable, $b_i$ is a boolean expression, $e_i$ is an expression (for $1 \leq i \leq n$),
and $s(v)$ is one of $v$, \texttt{init}($v$), or \texttt{next}($v$).
We denote by $\sbr{M}$ the FTS for this model~\cite{classenTR}.

Let $\Kk' \subseteq 2^{\Ff}$ be a set of configurations described by a feature expression $\psi'$, i.e.\ $\sbr{\psi'}=\Kk'$.
The projection $\pi_{\sbr{\psi'}}(\sbr{M})$ is obtained by adding
the \texttt{INVAR} constraint $\psi'$
to the model $M$, denoted as $M+\texttt{INVAR}(\psi')$.
Thus, $\pi_{\sbr{\psi'}}(\sbr{M}) = \sbr{M+\texttt{INVAR}(\psi')}$.
Another solution would be to add
the constraint $\psi'$ to each $b_i$ in the assignments to the state variables.

Let $(\alpha,\gamma)$ be a Galois connection from Section~\ref{sec:definition}.
The abstract $\alpha(M)^{may}$ and $\alpha(M)^{must}$ are obtained by
the following rewrites for assignments in $M$:
{\small
\[
\begin{array}{l}
%\scriptstyle{
\alpha \big( s(v) \!:=\! \texttt{case} \, b_1\!:\!e_1; \ldots b_n\!:\!e_n; \texttt{esac} \big)^{may} \!=\! %\\ \qquad
s(v) \!:=\! \texttt{case} \, \alpha^m(b_1)\!:\!e_1; \ldots \alpha^m(b_n)\!:\!e_n; \texttt{esac} \\
\alpha \big( s(v) \!:=\! \texttt{case} \ b_1\!:\!e_1; \ldots b_n\!:\!e_n; \texttt{esac} \big)^{must} \!=\! %\\ \qquad
s(v) \!:=\! \texttt{case} \ \widetilde{\alpha}(b_1)\!:\!e_1; \ldots \widetilde{\alpha}(b_n)\!:\!e_n; \texttt{esac}
%}
\end{array}
\]
}
The functions $\alpha^m$ and $\widetilde{\alpha}$ copy all basic boolean expressions other than feature expressions,
and recursively calls itself for all sub-expressions of compound expressions.
For $\joinasym(M)^{may}$, we have a single Boolean variable $rnd$ which is non-deterministically initialized.
Then, $\alpha^m(\psi)=rnd$ if $\alpha(\psi)=\true$.
%For $\joinasym$ and $\widetilde{\joinasym}$ we obtain a single \fsmv\ model where all
%features are removed, whereas for $\fignoreasym_{A}$ and $\widetilde{\fignoreasym_{A}}$
%we obtain an \fsmv\ model where only the feature $A$ is removed.
We have: $\alpha(\sbr{M})^{may} = \sbr{\alpha(M)^{may}}$ and $\alpha(\sbr{M})^{must} = \sbr{\alpha(M)^{must}}$.
For example, given the composed model $\Mmm$ in Fig.~\ref{fig:fsmv2},
the abstractions $\joinasym(\Mmm)^{may}$ and $\joinasym(\Mmm)^{must}$ are shown in Figs.~\ref{fig:abs:smv1} and \ref{fig:abs:smv2}, respectively.
Note that $\widetilde{\joinasym}(f.fA)=\false$, so the first branch of \texttt{case} statements
in $\Mmm$ is never taken in $\joinasym(\Mmm)^{must}$.

\begin{figure}[t]
\centering
\begin{minipage}[b]{.5\textwidth}
\centering
%\begin{figure}
$ \begin{array}{l}
\textcolor{gray}{1}~~ \texttt{MODULE \emph{main} } \\
\textcolor{gray}{2}~~ \quad \texttt{VAR}\ x:0..1; \, nA : 0..1; \, rnd : boolean; \\
\textcolor{gray}{3}~~ \quad \texttt{ASSIGN}\\
\textcolor{gray}{4}~~ \qquad \texttt{init}(x) := 0; \texttt{init}(nA) := 0; \\
\textcolor{gray}{5}~~ \qquad \texttt{init}(rnd) := \{\textit{TRUE,FALSE}\}; \\
% \texttt{next}(rnd) := rnd; \\
\textcolor{gray}{6}~~ \qquad \texttt{next}(x) := \texttt{case} \, rnd \, \& \, nA\!=\!0 : x+1; \\
\textcolor{gray}{7}~~ \qquad \qquad \qquad \qquad \quad \textit{TRUE : } x; \ \texttt{easc}; \\
\textcolor{gray}{8}~~ \qquad \texttt{next}(nA) := \texttt{case} \\
\textcolor{gray}{9}~~ \qquad \quad rnd \, \& \, nA\!=\!0 \, \& \, \texttt{next}(x)\!=\!x\!+\!1 : 1; \\
\textcolor{gray}{10}~~ \qquad \   \textit{TRUE : } nA; \ \texttt{easc};
\end{array} $
\vspace{-1.5mm}
\caption{$\joinasym(\Mmm)^{may}$}
\label{fig:abs:smv1}
%\end{figure}
\end{minipage}%
\begin{minipage}[b]{.5\textwidth}
\centering
%\begin{figure}
$ \begin{array}{l}
\textcolor{gray}{1}~~ \texttt{MODULE \emph{main} } \\
\textcolor{gray}{2}~~ \quad \texttt{VAR}\ x:0..1; \, nA : 0..1;\\
\textcolor{gray}{3}~~ \quad \texttt{ASSIGN}\\
\textcolor{gray}{4}~~ \qquad \texttt{init}(x) := 0; \texttt{init}(nA) := 0; \\
\textcolor{gray}{5}~~ \qquad \texttt{next}(x) := x; \\
\textcolor{gray}{6}~~ \qquad \texttt{next}(nA) := nA;
\end{array} $
\vspace{-1.5mm}
\caption{$\joinasym(\Mmm)^{must}$}
\label{fig:abs:smv2}
%\end{figure}
\end{minipage}
%\vspace{-4mm}
%\caption{Abstract \smv\ models.} \label{fig:abs:smv}
\end{figure}

\section{Evaluation} \label{sec:evaluation}

We now evaluate our abstraction-based verification technique.
First, we show how
variability abstractions can turn a previously infeasible analysis of variability
model into a feasible one.
Second, we show that
 instead of
verifying CTL properties  using
%a family-based model checker (e.g.,
the family-based version of \smv
\footnote{An extended version of \smv\, \cite{classen-model-checking-spls-icse11}
implements the family-based
algorithm for variational models obtained by composing the
basic model and all available features.},
we can use variability abstraction to obtain an abstract variability
model (with a low number of variants) that can be subsequently
model checked using %a single-system model checker (e.g.,
the standard version of \smv.

All experiments
were executed on a %LUbunutuVM
64-bit Intel$^\circledR$Core$^{TM}$ i7-4600U CPU running at 2.10 GHz with 8 GB
memory.
The implementation, benchmarks, and all results obtained from our experiments
are available from: \url{https://aleksdimovski.github.io/abstract-ctl.html}.
The reported performance numbers constitute the average runtime of five independent executions.
For each experiment, we report the
time needed to perform the verification task in seconds.
% and \textsc{Space} which is the number of
%explored states plus the number of re-explored states (this is
%equivalent to the number of transitions fired).
The BDD model checker \smv\, is run with the parameter \texttt{-df -dynamic}, which
ensures that the BDD package reorders the variables during verification in case the
BDD size grows beyond a certain threshold.
We consider two case studies:
a synthetic example to demonstrate specific characteristics of our approach,
and the \elevator\ system \cite{DBLP:journals/scp/PlathR01} which is
a standard benchmark in the SPLE community~\cite{classen-model-checking-spls-icse11,classenTR,DBLP:journals/sttt/ClassenCHLS12,sttt16}.

\paragraph{\textbf{Synthetic example.}}

As an experiment, we have tested limits of family-based model checking with extended \smv\ and
``brute-force'' single-system model checking with standard \smv\ (where all variants are verified one by one).
We have gradually added variability to the  variational model
in Fig.~\ref{fig:fsmv}.
This was done by adding optional features which increase the basic model's variable $x$ by
the number corresponding to the given feature.
For example, the \texttt{CHANGE} section for the second feature $B$ is:
$\texttt{IF} \, (nB=0) \, \texttt{THEN } \texttt{IMPOSE next}(x):=x\!+\!2; \, \texttt{next}(nB):=\texttt{next}(x)\!=\!x\!+\!2?1\!:\!nB$,
and the domain of $x$ is $0..3$.

We check the assertion $\forall \Diamond (x \geq 0)$.
For $|\Ff| = 25$ (for which $|\Kk| = 2^{25}$
variants, and the state space is $2^{32}$) the family-based \smv\ takes around 77 minutes to verify the assertion,
whereas for $|\Ff| = 26$ it has not finished the task within two hours.
The analysis time to check the assertion
using ``brute force'' with standard \smv\ ascends to almost three years for $|\Ff| = 25$.
On the other hand, if we apply the variability abstraction
$\joinasym$, we
are able to verify the same assertion by only one call to standard \smv\ on the
\emph{abstracted} model in 2.54 seconds for $|\Ff| = 25$ and in 2.99 seconds for $|\Ff| = 26$.
%, effectively eliminating the exponential blow up.

\paragraph{\textbf{\elevator.}}
The \elevator, designed by Plath and Ryan \cite{DBLP:journals/scp/PlathR01},
contains about 300 LOC and 9 independent features:
% (which are leaves in the tree organization of the feature model) <== NOT NECESSARILY!
\texttt{Antiprunk}, \texttt{Empty}, \texttt{Exec}, \texttt{OpenIfIdle},
\texttt{Overload}, \texttt{Park}, \texttt{QuickClose}, \texttt{Shuttle}, and
\texttt{TTFull}, thus yielding $2^9$ = 512 variants.
The elevator serves a number of floors (which is five in our case) such that there is a
single platform button on each floor which calls the elevator.
The elevator will always serve all requests in its current direction before it stops
and changes direction. When serving a floor, the elevator door opens and closes again.
The size of the \elevator\, model is $2^{28}$ states.
On the other hand, the sizes of $\joinasym(\elevator)^{may}$ and $\joinasym(\elevator)^{must}$
are $2^{20}$ and $2^{19}$ states, resp.

\begin{figure*}[t]
\begin{center}
\begin{tabular}{||c||cc||cc||c||}  \hhline{|t:=:t:==:t:==:t:=:t|}
~\emph{prop-}~ & \multicolumn{2}{c||}{\emph{family-based app.}}     & \multicolumn{2}{c||}{\emph{abstraction-based app.}} & \multicolumn{1}{c||}{\emph{improvement}} \\
~\emph{-erty}~  & ~~$|\Kk|$~~ & ~\textsc{Time}~~  & $|\alpha(\Kk)|$ & ~\textsc{Time}~~  & ~~\textsc{Time}~~      \\ \hhline{|:=::==::==::=:|}
$\Phi_1$ & 512 & 36.73 s & 2 & 2.59 s & \phantom{1}14 $\times$ \\ \hhline{||-||--||--||-||}
$\Phi_2$ & 512 & 35.89 s & 2 & 6.95 s & \phantom{1}5 $\times$ \\ \hhline{||-||--||--||-||}
$\Phi_3$ & 512 & 54.76 s & 1 & 1.67 s & \phantom{1}32 $\times$ \\ \hhline{||-||--||--||-||}
$\Phi_4$ & 512 & 2.65 s & 2 & 1.04 s & \phantom{1}2.5 $\times$ \\  \hhline{||-||--||--||-||}
$\Phi_5$ & 512 & 37.76 s & 2 & 2.62 s & \phantom{1}15 $\times$ \\ \hhline{|b:=:b:==:b:==:b:=:b|}
\end{tabular}
\vspace{-2mm}
\caption{Verification of \elevator\ properties using tailored abstractions.
We compare family-based approach vs. abstraction-based approach.
}\label{fig:casestudy:elevator}
\end{center}
\end{figure*}
%\vspace{-1mm}

We consider five properties.
The $\forall$CTL property
``$\Phi_1 = \forall \Box \,(floor=2 \land liftBut5.pressed \land direction=up \Rightarrow \forall [direction=up \, \U floor=5]$'' is that,
when the elevator is on the second floor with direction up and the button five is pressed, then the elevator
will go up until the fifth floor is reached.
This property is violated by variants for which \texttt{Overload} (the elevator will refuse to close its doors when it is overloaded) is satisfied.
Given sufficient knowledge of the system and the property, we can tailor an abstraction for
verifying this  property more effectively. We call standard \smv\, to check $\Phi_1$ on
two models $\joinasym(\pi_{\sbr{\texttt{Overload}}}(\elevator))^{may}$ and $\joinasym(\pi_{\sbr{\neg \texttt{Overload}}}(\elevator))^{may}$.
For the first abstracted projection we obtain an ``abstract'' counter-example violating $\Phi_1$,
whereas the second abstracted projection satisfies $\Phi_1$.
Similarly, we can verify that the $\forall$CTL property
``$\Phi_2 = \forall \Box \,(floor=2 \land direction=up \Rightarrow \forall \bigcirc (direction=up))$'' is satisfied
only by variants with enabled \texttt{Shuttle} (the lift will change direction at the first and last floor).
We can successfully verify $\Phi_2$ for $\joinasym(\pi_{\sbr{\texttt{Shuttle}}}(\elevator))^{may}$ and obtain a
counter-example for $\joinasym(\pi_{\sbr{\neg \texttt{Shuttle}}}(\elevator))^{may}$.
The $\exists$CTL property
``$\Phi_3 = (\texttt{OpenIfIdle} \land \neg \texttt{QuickClose}) \implies \exists \Diamond ( \exists \Box \,( door=open))$'' is that,
there exists an execution such that from some state on the door stays open.
We can invoke the standard \smv\, to verify that $\Phi_3$ holds for
$\joinasym(\pi_{\sbr{\texttt{OpenIfIdle} \land \neg \texttt{QuickClose}}}(\elevator))^{must}$.
The following two properties are neither in $\forall$CTL nor in $\exists$CTL.
The property
``$\Phi_4 = \forall \Box \,(floor=1 \land idle \land door=closed \implies \exists \Box (floor=1 \land door=closed))$'' is that,
for any execution globally if the elevator is on the first floor, idle, and its door is closed, then there is a continuation where
the elevator stays on the first floor with closed door.
The satisfaction of  $\Phi_4$ can be established by verifying it against both
$\joinasym(\elevator)^{may}$ and $\joinasym(\elevator)^{must}$ using two calls to standard \smv.
The property
``$\Phi_5 = \texttt{Park} \implies \forall \Box \,( floor=1 \land idle \implies \exists [idle \U floor=1] )$'' is
satisfied by all variants with enabled \texttt{Park} (when idle, the elevator returns to the first floor).
We can successfully  verify $\Phi_5$ by analyzing $\joinasym(\pi_{\sbr{\texttt{Park}}}(\elevator))^{may}$ and $\joinasym(\pi_{\sbr{\texttt{Park}}}(\elevator))^{must}$
using two calls to  standard $\smv$.
%The property $\varphi_4$ is in $\exists$CTL, so we can verify it against $\joinasym(\elevator)^{must}$ on standard $\smv$.
We can see in Fig.~\ref{fig:casestudy:elevator} that abstractions
achieve significant speed-ups between 2.5 and 32 times faster than the family-based approach.

\section{Related Work}\label{sec:related}

Recently, many family-based techniques that work on the level of variational systems have been proposed.
This includes family-based syntax checking \cite{DBLP:conf/oopsla/KastnerGREOB11,DBLP:conf/pldi/GazzilloG12},
family-based type checking \cite{DBLP:journals/tosem/KastnerATS12},
family-based static program analysis \cite{scp15,ecoop15,DBLP:conf/fm/DimovskiBW16},
family-based verification \cite{programming17,DBLP:journals/jlp/RheinTSLA16,DBLP:conf/kbse/Iosif-LazarADSS15}, etc.
In the context of family-based model checking,
one of the earliest attempts for modelling variational systems
is by using modal transition systems (MTSs) \cite{DBLP:conf/esop/LarsenNW07,Beek16}.
Subsequently,
Classen et al. present FTSs \cite{model-checking-spls} and
specifically designed family-based model checking algorithms for verifying FTSs
against LTL \cite{DBLP:journals/sttt/ClassenCHLS12}.
This approach is extended \cite{classen-model-checking-spls-icse11,classenTR} to enable verification of CTL properties using an
family-based version of \smv.
The work \cite{DBLP:conf/fase/BeekVW17} shows how modal $\mu$-calculus properties
of variational systems can be verified using a general-purpose model checker \textsf{mCRL2}.
In order to make this family-based approach more scalable, the works \cite{spin,sttt16} propose applying
conservative variability abstractions on FTSs
for deriving
abstract family-based model checking of LTL.
An automatic abstraction refinement procedure
  for family-based model checking is then proposed in \cite{fase17}, which works until a genuine counterexample is found or the
  property satisfaction is shown for all variants in the family.
The application of variability abstractions for verifying LTL and $\forall$CTL of real-time variational systems is described in  \cite{DBLP:conf/birthday/DimovskiW17}.
The works \cite{spin16,DBLP:journals/tcs/Dimovski18} present an approach for family-based software model checking
of \texttt{\#ifdef}-based (second-order) program families using symbolic game semantics models \cite{DTCS14,gandalf17}.

\section{Conclusion}

We have proposed conservative (over-approximating) and their dual (under-approximating) variability abstractions to derive abstract
family-based model checking that preserves the full CTL$^{\star}$.
The evaluation confirms that interesting properties can be efficiently verified in this way.
In this work, we assume that a suitable abstraction is manually generated before verification.
If we want to make the whole verification procedure automatic, we need to develop an abstraction and
refinement framework for CTL$^{\star}$ properties similar to the one in \cite{fase17} which is designed for LTL. 

%\vspace{-1mm}
%\section{Conclusion}\label{sec:conclusion}
%\input{conclusion.tex}

\bibliographystyle{splncs03}
\bibliography{ms}

\newpage
\appendix

\section{Proofs}\label{app:proofs}

\textbf{Lemma 1.}
Let $\psi  \in \FeatExp(\Ff)$, and $\Kk$ be a set of configurations over $\Ff$.
\begin{description}
\item[(i)] Let $k \in \Kk$ and $k \models \psi$. Then there exists $k' \in \alpha(\Kk)$, such that $k' \models \alpha(\psi)$.
\item[(ii)] Let $k' \in \alpha(\Kk)$ and $k' \models \widetilde{\alpha}(\psi)$. Then for all $k \in \Kk$ s.t. $\alpha(k)=k'$, it holds $k \models \psi$.
\end{description}
%\end{lemma}
\begin{proof}[Lemma\,\ref{lemma:1}]
By induction on the structure of $\alpha$.
\begin{description}
\item[(i)] The proof is similar to proof of Lemma 2 in \cite{sttt16}.
\item[(ii)]
  \begin{description}
  \item[Case $\joinasym$:]
By assumption, we have that $\Kk \neq \emptyset$, thus $\joinasym(\Kk)=\{\true\}$. 
We have $\joinasym(k)=\true$ for all $k \in \Kk$.
Since $\true \models \widetilde{\joinasym}(\psi)$, it follows that $\widetilde{\joinasym}(\psi)=\true$.
This is the case only if for all $k \in \Kk$, it holds $k \models \psi$.

\item[Case $\fignorea{A}$:]
By assumption, $k'=k[l_A \mapsto \true] \in \fignorea{A}(\Kk)$ and $k' \models \widetilde{\fignorea{A}}(\psi)$.
We have $\fignorea{A}(k)=k'$ for all $k \in \Kk$ s.t. $k[l_A \mapsto \true]=k'$. Since $k' \models \widetilde{\fignorea{A}}(\psi)$, 
we have $k[l_A \mapsto \true] \models \psi[l_A \mapsto \false]$.
Thus, we must have that $k \models \psi$ for all $k \in \Kk$ s.t. $\fignorea{A}(k)=k'$.
\end{description}
\end{description}
\qed
\end{proof}

\noindent
\textbf{Lemma 2.}
\begin{description}
\item[(i)] Let $k \in \Kk$ and $\rho \in \sbr{\pi_k(\Fff)}_{TS} \subseteq \sbr{\Fff}_{FTS}$. Then there exists $k' \in \alpha(\Kk)$, such that $\rho \in \sbr{\pi_{k'}(\alpha(\Fff))}^{may}_{MTS} \subseteq \sbr{\alpha(\Fff)}^{may}_{MFTS}$ is a may-execution in it.
\item[(ii)] Let $k' \in \alpha(\Kk)$ and $\rho \in \sbr{\pi_{k'}(\alpha(\Fff))}^{must}_{MTS} \subseteq \sbr{\alpha(\Fff)}^{must}_{MFTS}$ be a must-execution in it. Then for all $k \in \Kk$ s.t. $\alpha(k)=k'$, it holds $\rho \in \sbr{\pi_k(\Fff)}_{TS} \subseteq \sbr{\Fff}_{FTS}$.
\end{description}
%\end{lemma}
\begin{proof}[Lemma\,\ref{lemma:2}]
\begin{description}
\item[(i)]
Let $\rho = s_0 \lambda_1 s_1 \lambda_2 \ldots \in \sbr{\pi_k(\Fff)}_{TS}$.
This means that for all transitions in $\rho$, $t_i=s_i \stackrel{\lambda_{i+1}}{\longrightarrow} s_{i+1}$,
we have that $k \models \delta(t_i)$ for all $i \geq 0$.
By Lemma~\ref{lemma:1}(i), we have that there exists $k' \in \alpha(\Kk)$,
such that $k' \models \alpha(\delta(t_i))$, i.e.\ $k' \models \delta^{may}(t_i)$, for all $i \geq 0$.
Hence, we have $\rho \in \sbr{\pi_{k'}(\alpha(\Fff))}^{may}_{MTS}$.
\item[(ii)]
Let $\rho = s_0 \lambda_1 s_1 \lambda_2 \ldots \in \sbr{\pi_{k'}(\alpha(\Fff))}^{must}_{MTS}$.
This means that for all transitions in $\rho$, $t_i=s_i \stackrel{\lambda_{i+1}}{\longrightarrow} s_{i+1}$,
we have that $k' \models \widetilde{\alpha}(\delta(t_i))$, i.e.\ $k' \models \delta^{must}(t_i)$, for all $i \geq 0$.
By Lemma~\ref{lemma:1}(ii), we have that for all $k \in \Kk$ s.t. $\alpha(k)=k'$, it holds $k \models \delta(t_i)$ for all $i \geq 0$.
Hence, we have $\rho \in \sbr{\pi_k(\Fff)}_{TS}$ for all $k \in \Kk$ s.t. $\alpha(k)=k'$.
\end{description}
\qed
\end{proof}
\noindent
\textbf{Theorem 1.}[Preservation of CTL$^{\star}$]
%For every $\Phi \in CTL^{\star}$,
$\alpha(\Fff) \models \Phi \ \implies \ \Fff \models \Phi$.
%\end{theorem}
\begin{proof}[Theorem\,\ref{theorem:sound}]
We prove the most difficult case [CTL$^\star$].
By induction on the structure of $\Phi$.
We prove for state formulae $\Phi$ that if $\alpha(\Fff) \models \Phi$ (i.e.\ $\pi_{k'}(\alpha(\Fff)) \models \Phi$ for all $k' \in \alpha(\Kk)$),
then $\Fff \models \Phi$ (i.e.\ for all $k \in \Kk$, $\pi_{k}(\Fff) \models \Phi$).
All cases except $\forall$ and $\exists$ quantifiers are straightforward.

For $\Phi=\forall \phi$, we proceed by contraposition.
Assume $\Fff \not\models \forall \phi$. Then, there exists a configuration $k \in \Kk$
and an execution $\rho \in \sbr{\pi_k(\Fff)}_{TS}$ such that $\rho \not\models \phi$, i.e.\ $\rho \models \neg\phi$.
By Lemma~\ref{lemma:2}(i), we have that there exists $k' \in \alpha(\Kk)$,
such that $\rho \in \sbr{\pi_{k'}(\alpha(\Fff))}^{may}_{MTS}$, and so
$\alpha(\Fff) \not\models \forall \phi$.

For $\Phi=\exists \phi$.
Assume $\alpha(\Fff) \models \exists \phi$. Then, for all configurations $k' \in \alpha(\Kk)$, 
we have $\pi_{k'}(\alpha(\Fff)) \models \exists \phi$. 
This means that there exists an execution $\rho \in \sbr{\pi_{k'}(\alpha(\Fff))}^{must}_{MTS}$ such that $\rho \models \phi$.
By Lemma~\ref{lemma:2}(ii), we have that for all $k \in \Kk$ s.t. $\alpha(k)=k'$,
we have $\rho \in \sbr{\pi_{k}(\Fff)}_{TS}$, and so $\pi_k(\Fff) \models \exists \phi$. 
Since for any $k \in \Kk$, there exists some $k' \in \alpha(\Kk)$ s.t. $\alpha(k)=k'$, we have that 
$\pi_k(\Fff) \models \exists \phi$ for all $k \in \Kk$, and so $\Fff \models \exists \phi$. \qed
\end{proof}

\noindent
\textbf{Theorem 2.}
For every  $\Phi \in CTL^{\star}$ and MFTS $\Mff$, we have:
\[
\Mff \models \Phi = \begin{cases} \true & \textrm{if  } \big( \Mff^{may} \models \Phi \, \land \, \Mff^{must} \models \Phi \big) \\
\false & \textrm{if  } \big( \Mff^{may} \not\models \Phi \, \lor \, \Mff^{must} \not\models \Phi \big) \end{cases}
\]
%\end{theorem}
\begin{proof}[Theorem\,\ref{theorem:mfts}]
By induction on the structure of $\Phi$. See Appendix~\ref{app:proofs}.
All cases except $\forall$ and $\exists$ quantifiers are straightforward.

For $\Phi=\forall \phi$. Consider the first case, when $\Mff \models \Phi = \true$.
Assume $\Mff^{may} \models \forall \phi$. That is, for any may-execution
$\rho$ of $\Mff$ we have $\rho \models \phi$. By Definition \ref{def:ctl} (3'), we have $\Mff \models \Phi$.
Consider the second case, when $\Mff \models \Phi = \false$.
Assume $\Mff^{may} \not\models \forall \phi$. That is, there exists a may-execution
$\rho$ of $\Mff$ such that $\rho \models \phi$. By Definition \ref{def:ctl} (3'), we have $\Mff \not\models \Phi$.
Assume $\Mff^{must} \not\models \forall \phi$. That is, there exists a must-execution
$\rho$ of $\Mff$ such that $\rho \not\models \phi$.
But $\rho$ ia also a may-execution, so
by Definition \ref{def:ctl} (3'), we have $\Mff \not\models \Phi$.

For $\Phi=\exists \phi$.
Consider the first case, when $\Mff \models \Phi = \true$.
Assume $\Mff^{must} \models \exists \phi$. That is, there exists a must-execution
$\rho$ of $\Mff$ such that $\rho \models \phi$. By Definition \ref{def:ctl} (3'), we have $\Mff \models \Phi$.
Consider the second case, when $\Mff \models \Phi = \false$.
Assume $\Mff^{may} \not\models \exists \phi$. That is, for all may-executions
$\rho$ of $\Mff$ we have $\rho \not\models \phi$. Since all must-executions are also may-executions,
we have that all must-executions do not satisfy $\phi$.
By Definition \ref{def:ctl} (3'), we have $\Mff \not\models \Phi$.
Assume $\Mff^{must} \not\models \exists \phi$. That is, for all must-executions
$\rho$ of $\Mff$ we have $\rho \not\models \phi$.
By Definition \ref{def:ctl} (3'), we have $\Mff \not\models \Phi$.
\end{proof}

\newpage 
\section{Figures}\label{app:figures}

\begin{figure}
\centering
\begin{tikzpicture}[->,>=stealth',shorten >=0.9pt,auto,node distance=1.45cm,
                    semithick]
  \tikzstyle{every state}=[minimum size=.2pt,initial text={{\scriptsize start}}]

  \node[initial below,state] (A)     {\tiny 1};
  \node[state]         (B) [right=7.5mm of A] {\tiny 2};
  \node[state]         (C) [right=11mm of B] {\tiny 3};
  \node[state]         (D) [above right=8mm of A]  {\tiny 4};
  \node[state]         (E) [above right=11mm of C]  {\tiny 5};
  \node[state]         (F) [below right=4mm of C]  {\tiny 6};
  \node[state]         (G) [right=20mm of C]  {\tiny 7};
  \node[state]         (H) [right=7.5mm of G]  {\tiny 8};
  %\node[state]         (I) [right=9.5mm of H]  {\tiny 9};

  \path[font=\tiny] (A) edge              node {$\textit{pay}$} (B)
        (B) edge              node {$\textit{change}$} (C)
        (G) edge   node {$\textit{open}$} (H)
        (H) edge [in=-70,out=-25]  node[above right,sloped] {$\textit{take}$} (A);

  \path[font=\tiny]      (C) edge [bend left] node[above,sloped] {$\textit{soda}$} (E)
        (E) edge [bend left] node[above,sloped] {$\textit{serveSoda}$} (G);

  \path[draw,dashed,font=\tiny]      (A) edge [bend right] node[below,sloped] {{$\textit{free}$}} (C)
        (G) edge [bend left=56,in=128,out=128]  node[above,sloped] {{$\textit{take}$}} (A);

  \path[draw=brown,fill=brown,font=\tiny]
        (G) edge [in=68,out=128] node[below,sloped] {{\color{brown}$\textit{take} / c$}} (C)
        (C) edge [bend right] node[above,sloped] {{\color{brown}$\textit{cancel} / c$}} (D)
        (D) edge [bend right] node[above,sloped] {{\color{brown}$\textit{return} /\! c$}} (A);

  \path[draw, dashed, font=\tiny]
        (C) edge [bend right] node[below,sloped] {$\textit{tea} $} (F)
        (F) edge [bend right,in=180] node[above left] {$\textit{serveTea}$} (G);

\end{tikzpicture}
\vspace{-10mm}
\caption{$\fignorea{\{\ft,\ff\}}(\pi_{\sbr{\fv \,\land\, \fs}}(\textsc{VendingMachine}))$. For clarity, we omit to write the presence condition \true\, in transitions. }
\label{fig:abs4}
\end{figure}

\end{document}